\documentclass[aps,prb,fleqn,12pt,superscriptaddress]{revtex4}
\usepackage{epsfig,color}
\usepackage{graphicx}
\usepackage{pdfpages}
\usepackage{braket}
\usepackage{amssymb,amsmath}
\usepackage{hyperref}
\begin{document}
\title{Orbital entangled antiferromagnetoc order and spin-orbit-distortion exciton in {\boldmath $\rm Sr_2VO_4$}} 
\author{Shubhajyoti Mohapatra}
\affiliation{Department of Physics, Indian Institute of Technology, Kanpur - 208016, India}
\author{Dheeraj Kumar Singh}
\affiliation{Thapar Institute of Engineering and Technology, Patiala - 147004, India}
\author{Rajyavardhan Ray}
\affiliation{Department of Physics, Birla Institute of Technology Mesra, Ranchi - 835215, India}
\author{Sayandip Ghosh}
\affiliation{D\'{e}partement de physique, Institut quantique and Regroupement Qu\'eb\'ecois sur les Mat\'eriaux de Pointe, Universit\'{e} de Sherbrooke, Sherbrooke, Qu\'{e}bec J1K 2R1, Canada}
\author{Avinash Singh}
\email{avinas@iitk.ac.in}
\affiliation{Department of Physics, Indian Institute of Technology, Kanpur - 208016, India}
\date{\today} 
\begin{abstract}
With electron filling $n=1$ in the $\rm Sr_2VO_4$ compound, the octahedrally coordinated $t_{\rm 2g}$ orbitals are strongly active due to tetragonal distortion induced crystal field tuning by external agent such as pressure. Considering the full range of crystal field induced tetragonal splitting in a realistic three-orbital model, collective spin-orbital excitations are investigated using the generalized self consistent and fluctuation approach. The variety of self consistent states obtained including orbital entangled ferromagnetic and antiferromagnetic orders reflects the rich spin-orbital physics resulting from the interplay between the band, spin-orbit coupling, crystal field, and Coulomb interaction terms. The behavior of the calculated energy scales of collective excitations with crystal field is consistent with that of the transition temperatures with pressure as obtained from susceptibility and resistivity anomalies in high-pressure studies. 
\end{abstract}
\maketitle
\newpage

\section{Introduction}

The origin and nature of the different phases observed in the transition metal oxide $\rm Sr_2VO_4$ has been a recurrent theme despite early synthesis and significant effort made afterwards in both theoretical and experimental directions.\cite{cyrot_JSSC_1990,rey_JSSC_1990} $\rm Sr_2VO_4$ was considered a promising candidate for unconventional superconductivity,\cite{nozaki_PRB_1991} mainly because the layered crystal structure consisting of VO$_6$ octahedra having V$^{4+}$ ions with 3$d^1$ configuration bears a remarkable similarity (in view of electron-hole symmetry) to the parent compound $\rm La_2CuO_4$ of high-$T_c$ cuprates having Cu$^{2+}$ ions with 3$d^9$ configuration in CuO$_2$ layers.\cite{lee_RMP_2006} While superconductivity remains elusive, the origin of phases including the magnetically ordered one are not well understood. Unlike the cuprates, the complexity in understanding the vanadate system stems from factors including the active orbital degree of freedom, non-negligible spin-orbit coupling (SOC), Coulomb interaction induced orbital mixing terms and electron-lattice coupling. 

$\rm Sr_2VO_4$ crystalizes into $\rm K_2NiF_4$-type structure found in compounds such as $\rm Sr_2CrO_4$, $\rm La_2CuO_4$ etc.\cite{cyrot_JSSC_1990} The $\rm VO_6$ octahedra are elongated along the $c$-axis. The octahedral-crystal field leads to partial lifting of the five-fold degenerancy of 3$d$ levels, which separate into two different sets of degenerate $e_g$ and $t_{2g}$ levels. The elongation of VO$_6$ along $c$-axis further removes the three-fold degenerancy so that $t_{2g}$ consists of low-lying doubly degenerate levels for the $d_{xz}$ and $d_{yz}$ orbitals and higher energy level for the $d_{xy}$ orbital. The single 3$d$ electron of V$^{4+}$ may occupy either of $d_{xz}$ or $d_{yz}$ orbitals, a picture expected in the absence of SOC. However, the orbital moments are unquenched and the SOC parameter is not negligibly small.\cite{abragam_book_1970} Therefore, the $t_{2g}$ manifold splits into $J = 1/2$ doublet and $J = 3/2$ quartet instead. Thus, single 3$d$ electron of V$^{4+}$ electron may occupy the low-lying $J = 3/2$ quartet states $|yz,\sigma\rangle \pm i |xz,\sigma\rangle$ with $\sigma=\uparrow,\downarrow$.\cite{jackeli_PRL_2009}

$\rm Sr_2VO_4$ exhibits several phase transitions both as a function of temperature as well as pressure. It undergoes three phase transitions at temperatures $T_0 \sim 10$K, $T_1 \sim 100$K and $T_2 \sim 130$K, as indicated by the magnetic susceptibility, electrical resistivity, and heat capacity measurements near ambient pressure.\cite{cyrot_JSSC_1990,zhou_PRL_2007,yamauchi_PRB_2015,yamauchi_PRL_2019} There is a structural transition from the tetragonal to the intermediate phase at $T_2 \sim 130$K marked by the onset of short-range orbital order. Whereas the phase transition at $T_1 \sim 100$K involves a structural change from the intermediate to the tetragonal phase accompanied with an abrupt increase in $c/a$ ratio. Since the magnetic susceptibility also drops below $T_1$, it was suggested that the transition was to an antiferromagnetic state with orbital order while the N\'{e}el temperature was found to be dependent on the sample quality. Although long-range magnetic order could not be confirmed in early neutron-scattering experiments,\cite{cyrot_JSSC_1990} there are several evidences based on magnetic susceptibility, electrical resistivity, and muon spin rotation measurements,\cite{cyrot_JSSC_1990,yamauchi_PRL_2019,sugiyama_PRB_2014} which imply that the phase transition to a magnetic insulator phase occurs at $T_0 \sim 10$K. However, the nature of magnetic order in the insulating state remains highly controversial. 

Inelastic neutron-scattering study of $\rm Sr_2VO_4$ shows two non-trivial excitation modes around 120 meV at low (4K) temperature.\cite{zhou_PRB_2010} The presence of finite intensity for neutron scattering is described to be an essential signature supporting the existence of the entangled orbital state and finite spin-orbit coupling. These 120 meV excitations persist up to 400K with gradually decreasing intensity and revealed a transition from a high-temperature orbital liquid phase (strong orbital fluctuation) to a low-temperature entangled orbital phase. The ordering temperature for the orbital liquid to the orbital-order phase transition is also found to be suppressed with increasing pressure due to enhanced orbital fluctuations. An orbital ordering transition was also found at $\sim$ 120K by analyzing the integrated area of the excitation peak. 

Several theoretical approaches have been adopted in order to examine the low-temperature phases observed in $\rm Sr_2VO_4$. The earliest band-structure calculations\cite{pickett_Physica_1991} indicated a magnetic instability that was confirmed by studies based on local density approximation (LDA + $U$) and Hartree-Fock approximation,\cite{imai_JPSJ_2006} and a ferromagnetic insulating ground state was found in both methods. On the other hand, a relativistic density-functional theory combined with an extended spin-1/2 Heisenberg model suggested the formation of spin liquid at low temperature.\cite{kim_PRB_2017} Studies based on first-principle calculation combined with path-integral renormalization group, where SOC was not incorporated, have examined the stabilization of different types of spin orderings such as ferromagnetic, double stripe, parquet, and have ruled these out in favour of an antiferromagnetic ground state with orbital order.\cite{imai_JPSJ_2006,imai_PRL_2005} Studies based on the Kugel-Khomskii model found that an entangled spin-orbital or magneticaly hidden octupolar order may be stabilized.\cite{jackeli_PRL_2009,eremin_PRB_2011} These studies lead to a diverging scenario with regard to the magnetic nature of low temperature ground state.

In this work, we will therefore investigate spin-orbital ordering and excitations in $\rm Sr_2VO_4$ using a unified approach which treats on an equal footing the various physical elements (SOC, Coulomb interaction, crystal field, and hopping terms) within a realistic microscopic model, and also allows for different orbital and magnetic orders (staggered, entangled, ferromagnetic, antiferromagnetic) as well as different spin-orbital fluctuations (magnon, orbiton, spin-orbiton). 

The structure of this paper is as below. The three-orbital model including hopping, crystal field, Coulomb interaction, and SOC terms is introduced in Sec. II. The generalized self-consistent plus fluctuation approach is briefly reviewed in Sec. III, and results of the calculation are presented in Sec. IV for the staggered orbital order and orbital entangled FM and AFM orders. The collective excitation modes calculated from the generalized fluctuation propagator over a broad crystal field range are discussed in Section V. The extremely low-energy orbiton excitation modes are identified as spin-orbit-distortion excitons in Sec. VI, and the behavior of these excitation energies with crystal field is discussed in Sec. VII and compared with that of the measured transition temperatures in high-pressure studies. Finally, some conclusions are presented in Sec. VIII. 

\section{Three-orbital model with SOC and Coulomb interactions}
In the three-orbital ($\mu=yz,xz,xy$), two-spin ($\sigma=\uparrow,\downarrow$) basis defined with respect to a common spin-orbital coordinate axes along the planar V-O-V directions, we consider the Hamiltonian ${\cal H} = {\cal H}_{\rm band} + {\cal H}_{\rm cf} + {\cal H}_{\rm int} + {\cal H}_{\rm SOC}$ within the $t_{\rm 2g}$ manifold. The band and crystal field terms have been discussed earlier,\cite{mohapatra_JPCM_2020} and are briefly summarized below.

The first, second, and third neighbor hopping terms for the $xy$ orbital are represented by $t_1$, $t_2$, $t_3$, respectively. For the $yz$ ($xz$) orbital, $t_4$ and $t_5$ are the nearest-neighbor (NN) hopping terms in $y$ $(x)$ and $x$ $(y)$ directions, respectively, corresponding to $\pi$ and $\delta$ orbital overlaps. The $xy$ orbital energy offset $\epsilon_{xy}$ (relative to the degenerate $yz/xz$ orbitals) represents the effective crystal field splitting, including the octahedral distortion (elongation/flattening) effect. We have taken hopping parameter values: ($t_1$, $t_2$, $t_3$, $t_4$, $t_5$)=$(-1.0, 0.3, 0, -1.0, 0.2)$ unless otherwise mentioned, and considered the range $+1\gtrsim \epsilon_{xy} \gtrsim -1$, all in units of the realistic hopping energy scale $|t_1|$=250 meV as obtained for the $\rm Sr_2CrO_4$ compound.\cite{pandey_PRB_2021} As there is no experimental evidence for octahedral rotation/tilting in $\rm Sr_2VO_4$, the orbital mixing hopping terms $t_{m1,m2,m3}$ have been set to zero.  

For the on-site Coulomb interaction terms in the $t_{2g}$ basis ($\mu,\nu=yz,xz,xy$), we consider:
\begin{eqnarray}
{\cal H}_{\rm int} &=& U\sum_{i,\mu}{n_{i\mu\uparrow}n_{i\mu\downarrow}} + U^\prime \sum_{i,\mu < \nu,\sigma} {n_{i\mu\sigma} n_{i\nu\overline{\sigma}}} + (U^\prime - J_{\mathrm H}) \sum_{i,\mu < \nu,\sigma}{n_{i\mu\sigma}n_{i\nu\sigma}} \nonumber\\ 
&+& J_{\mathrm H} \sum_{i,\mu \ne \nu} {a_{i \mu \uparrow}^{\dagger}a_{i \nu\downarrow}^{\dagger}a_{i \mu \downarrow} a_{i \nu \uparrow}} + J_{\mathrm P} \sum_{i,\mu \ne \nu} {a_{i \mu \uparrow}^{\dagger} a_{i \mu\downarrow}^{\dagger}a_{i \nu \downarrow} a_{i \nu \uparrow}} \nonumber\\ 
&=& U\sum_{i,\mu}{n_{i\mu\uparrow}n_{i\mu\downarrow}} + U^{\prime \prime}\sum_{i,\mu<\nu} n_{i\mu} n_{i\nu} - 2J_{\mathrm H} \sum_{i,\mu<\nu} {\bf S}_{i\mu}.{\bf S}_{i\nu} 
+J_{\mathrm P} \sum_{i,\mu \ne \nu} a_{i \mu \uparrow}^{\dagger} a_{i \mu\downarrow}^{\dagger}a_{i \nu \downarrow} a_{i \nu \uparrow} 
\label{h_int}
\end{eqnarray} 
including the intra-orbital $(U)$ and inter-orbital $(U')$ density interaction terms, the Hund's coupling term $(J_{\rm H})$, and the pair hopping interaction term $(J_{\rm P}=J_{\rm H})$, with $U^{\prime\prime} \equiv U^\prime-J_{\rm H}/2=U-5J_{\rm H}/2$ from the spherical symmetry condition $U^\prime=U-2J_{\mathrm H}$. Here $a_{i\mu\sigma}^{\dagger}$ and $a_{i\mu \sigma}$ are the electron creation and annihilation operators for site $i$, orbital $\mu$, spin $\sigma=\uparrow ,\downarrow$. The density operator $n_{i\mu\sigma}=a_{i\mu\sigma}^\dagger a_{i\mu\sigma}$, total density operator $n_{i\mu}=n_{i\mu\uparrow}+n_{i\mu\downarrow}=\psi_{i\mu}^\dagger \psi_{i\mu}$, and spin density operator ${\bf S}_{i\mu} = \psi_{i\mu}^\dagger ${\boldmath $\sigma$}$ \psi_{i\mu}$ in terms of the electron field operator $\psi_{i\mu}^\dagger=(a_{i\mu\uparrow}^{\dagger} \; a_{i\mu\downarrow}^{\dagger})$. All interaction terms above are SU(2) invariant and thus possess spin rotation symmetry.  

Finally, we consider the spin-space representation:
\begin{eqnarray} 
{\cal H}_{\rm SOC} (i) & = & -\lambda {\bf L}.{\bf S} = -\lambda (L_z S_z + L_x S_x + L_y S_y) \nonumber \\ 
&=& \left [ \begin{pmatrix} \psi_{yz \uparrow}^\dagger & \psi_{yz \downarrow}^\dagger \end{pmatrix} \begin{pmatrix} i \sigma_z \lambda /2 \end{pmatrix} 
\begin{pmatrix} \psi_{xz \uparrow} \\ \psi_{xz \downarrow} \end{pmatrix}
+ \begin{pmatrix} \psi_{xz \uparrow}^\dagger & \psi_{xz \downarrow}^\dagger \end{pmatrix} \begin{pmatrix} i \sigma_x \lambda /2 \end{pmatrix} 
\begin{pmatrix} \psi_{xy \uparrow} \\ \psi_{xy \downarrow} \end{pmatrix} \right . \nonumber \\
& + & \left . \begin{pmatrix} \psi_{xy \uparrow}^\dagger & \psi_{xy \downarrow}^\dagger \end{pmatrix} \begin{pmatrix} i \sigma_y \lambda /2 \end{pmatrix} 
\begin{pmatrix} \psi_{yz \uparrow} \\ \psi_{yz \downarrow} \end{pmatrix} \right ] + {\rm H.c.}
\label{soc}
\end{eqnarray}
for the bare spin-orbit coupling term, which explicitly breaks SU(2) spin rotation symmetry and therefore generates anisotropic magnetic interactions from its interplay with other Hamiltonian terms. In the following, we will consider bare SOC value $\lambda=0.2$ in the energy scale unit $|t_1|$=250 meV, which yields the realistic value $\lambda=50$ meV for $3d$ elements.\cite{radwanski_APP_2000,stohr_2006} 

\section{Generalized self consistent + fluctuation approach}
The generalized self consistent approach including all orbital diagonal and off-diagonal spin and charge condensates has been applied recently to the $\rm NaOsO_3$, $\rm Ca_2RuO_4$, and $\rm Sr_2IrO_4$ compounds,\cite{mohapatra_JPCM_2020,mohapatra_JPCM_2021,mohapatra_JMMM_2021} illustrating the rich interplay between different physical elements. The coupling of orbital moments to weak orbital fields and the interaction-induced SOC renormalization highlight the role of orbital off-diagonal condensates on the emergent orbital and spin-orbital physics. This approach has recently been extended to the chromate compound $\rm Sr_2CrO_4$,\cite{chromate_2021} where a SOC induced staggered-to-entangled orbital order transition was found in the reversed crystal field regime corresponding to ambient pressure. 

Resulting from orbital off-diagonal (OOD) spin and charge condensates, the additional contributions of the Coulomb interaction terms (Eq. \ref{h_int}) included in the generalized self consistent approach are:
\begin{equation}
[{\cal H}_{\rm int}^{\rm HF}]_{\rm OOD} = \sum_{i,\mu < \nu} \psi_{i\mu}^{\dagger} \left [
-\makebox{\boldmath $\sigma . \Delta$}_{i\mu\nu} + {\cal E}_{i\mu\nu} {\bf 1} \right ] \psi_{i\nu} + {\rm H.c.}
\label{h_hf_od} 
\end{equation}  
where the orbital off-diagonal spin and charge fields are self-consistently determined from:
\begin{eqnarray}
\makebox{\boldmath $\Delta$}_{i\mu\nu} &=& \left (\frac{U''}{2} + \frac{J_{\rm H}}{4} \right ) \langle \makebox{\boldmath $\sigma$}_{i\nu\mu} \rangle + \left (\frac{J_{\rm P}}{2} \right ) \langle \makebox{\boldmath $\sigma$}_{i\mu\nu} \rangle \nonumber \\
{\cal E}_{i\mu\nu} &=& \left (-\frac{U''}{2} + \frac{3J_{\rm H}}{4} \right ) \langle n_{i\nu\mu} \rangle + \left (\frac{J_{\rm P}}{2}\right ) \langle n_{i\mu\nu} \rangle  
\label{sc_od}
\end{eqnarray}
in terms of the corresponding condensates $\langle \makebox{\boldmath $\sigma$}_{i\mu\nu}\rangle \equiv \langle \psi_{i\mu}^{\dagger} \makebox{\boldmath $\sigma$} \psi_{i\nu} \rangle$ and $\langle n_{i\mu\nu} \rangle \equiv \langle \psi_{i\mu}^{\dagger} {\bf 1} \psi_{i\nu} \rangle$. The orbital mixing terms above explicitly preserve spin rotation symmetry, and are generally finite due to orbital mixing induced by SOC or octahedral tilting/rotation. In the following, we will see that these terms can also be generated spontaneously. 

Besides the spin magnetic moments in the AFM state, the orbital magnetic moments and Coulomb renormalized SOC values are also studied for different $\epsilon_{xy}$ values. The orbital moments and Coulomb interaction induced SOC renormalization were calculated from the orbital off-diagonal charge and spin condensates:
\begin{eqnarray}
\langle L_\alpha \rangle & = & -i \left [ \langle \psi_\mu^\dagger \psi_\nu\rangle - \langle \psi_\mu^\dagger \psi_\nu\rangle^* \right ] = 2\ {\rm Im}\langle \psi_\mu^\dagger \psi_\nu\rangle \nonumber \\
\lambda^{\rm int}_\alpha & = & (U'' - J_{\rm H}/2) {\rm Im}\langle \psi_\mu^\dagger \sigma_\alpha \psi_\nu\rangle
\label{phys_quan}
\end{eqnarray}
where the orbital pair ($\mu,\nu$) corresponds to the component $\alpha=x,y,z$. The last equation yields the Coulomb renormalized SOC values $\lambda_\alpha = \lambda + \lambda_\alpha^{\rm int}$ where $\lambda$ is the bare SOC value. Although the SOC-like $\lambda^{\rm int}_\alpha L_\alpha S_\alpha$ terms are a subset of Eq. (\ref{h_hf_od}) which explicitly preserves spin rotation symmetry,\cite{mohapatra_JMMM_2021} effectively enhanced SOC and magnetic anisotropy effect is seen in the generalized self consistent calculation due to the Coulomb orbital mixing terms.\cite{chromate_2021}


Since all generalized spin $\langle \psi_\mu ^\dagger \makebox{\boldmath $\sigma$}\psi_\nu \rangle$ and charge $\langle \psi_\mu ^\dagger \psi_\nu \rangle$ condensates are included in the self consistent approach, the fluctuation propagator must also be defined in terms of the generalized operators. We therefore consider the time-ordered generalized fluctuation propagator:
\begin{equation}
[\chi({\bf q},\omega)] = \int dt \sum_i e^{i\omega(t-t')} 
e^{-i{\bf q}.({\bf r}_i -{\bf r}_j)} 
\times \langle \Psi_0 | T [\sigma_{\mu\nu}^\alpha (i,t) \sigma_{\mu'\nu'}^{\alpha'} (j,t')] |\Psi_0 \rangle 
\end{equation}
in the self-consistent AFM ground state $|\Psi_0 \rangle$. The generalized spin-charge operators at lattice sites $i,j$ are defined as $\sigma_{\mu\nu}^\alpha = \psi_\mu ^\dagger \sigma^\alpha \psi_\nu$, which include both orbital diagonal ($\mu=\nu$) and off-diagonal ($\mu\ne\nu$) cases, as well as the spin ($\alpha=x,y,z$) and charge ($\alpha=c$) operators, with $\sigma^\alpha$ defined as Pauli matrices for $\alpha=x,y,z$ and unit matrix for $\alpha=c$. 

The generalized fluctuation propagator in the random phase approximation (RPA) was investigated recently for several $4d$ and $5d$ compounds with electron fillings $n=3,4,5$ in the $t_{\rm 2g}$ sector.\cite{mohapatra_JPCM_2021} Collective excitations have also been investigated recently for the $n=2$ $\rm Sr_2CrO_4$ compound,\cite{chromate_2021} which is of particular interest due to the active $yz/xz$ orbital degree of freedom since $n_{yz}+n_{xz}\approx 1$ in the reversed crystal field ($\epsilon_{xy}\sim -1$) regime where $n_{xy}\approx 1$. Since the generalized spin and charge operators $\psi_\mu ^\dagger \sigma^\alpha \psi_\nu$ include spin ($\mu=\nu$, $\alpha=x,y,z$), orbital ($\mu\ne\nu$, $\alpha=c$), and spin-orbital ($\mu\ne\nu$, $\alpha=x,y,z$) cases, the spectral function of the fluctuation propagator:
\begin{equation}
{\rm A}_{\bf q}(\omega) = \frac{1}{\pi} {\rm Im \; Tr} [\chi({\bf q}, \omega)]_{\rm RPA}
\label{spectral}
\end{equation}
provides information about the collective excitations (magnon, orbiton, and spin-orbiton), where the character is determined from the basis resolved contributions in the composite $\mu\nu\alpha$ basis. Orbiton and spin-orbiton modes correspond to same-spin and spin-flip particle-hole excitations, respectively, involving different orbitals. 

The generalized self consistent approach allows for staggered spin and orbital ordering, and also for orbital entangled ferromagnetic (FM) and antiferromagnetic (AFM) orders. In our self consistent calculation, we find that for fixed SOC the staggered orbital order (or equivalently antiferro-orbital (AFO) order) is unstable towards entangled FM ($z$) order or entangled AFM (planar) order depending on the $U$ and $J_{\rm H}$ values. The phase boundary between the entangled FM and AFM orders is therefore also investigated below covering a broad range of $U$ and $J_{\rm H}$ values. We will initially consider two specific cases: (i) $U=8$, $J_{\rm H}=U/6$ and (ii) $U=12$, $J_{\rm H}=U/10$ corresponding to different sides of the phase boundary. 

\section{Orbital entangled FM and AFM orders}

\begin{figure}
\vspace*{0mm}
\hspace*{0mm}
\psfig{figure=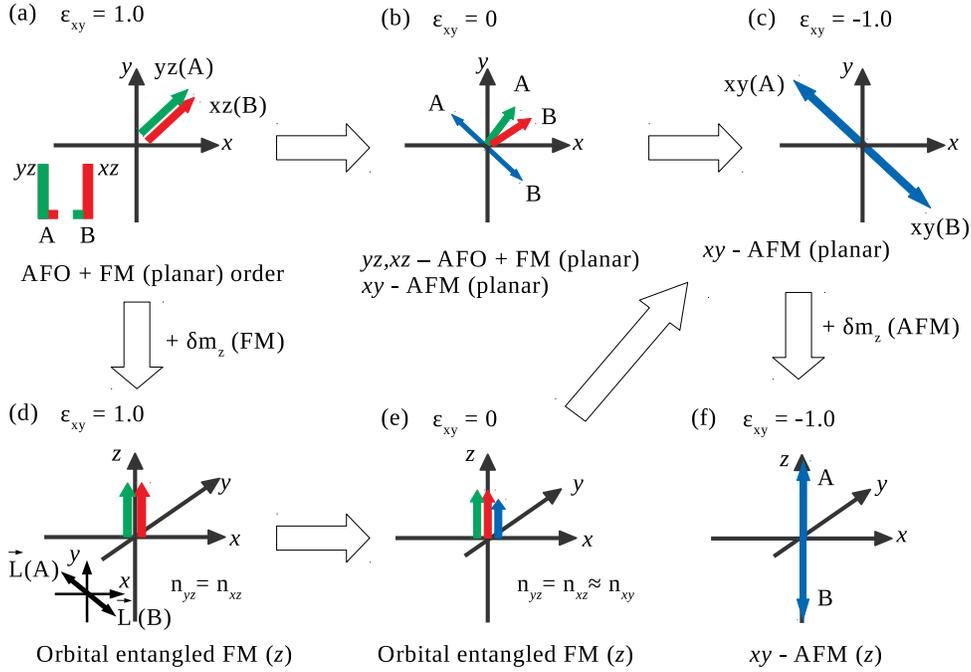,angle=0,width=130mm} \vspace{-10mm}
\caption{Evolution of the self-consistent spin-orbital order with the crystal field term $\epsilon_{xy}$. The AFO + FM (planar) order (a) at $\epsilon_{xy} \sim +1$ smoothly evolves to the $xy$ AFM (planar) order (c) at $\epsilon_{xy} \sim -1$ and then to $xy$ AFM ($z$) order (e) when small magnetization perturbation $\delta m_z$ is included. The staggered orbital order (a) is unstable towards orbital entangled FM ($z$) order (d,e) in which the planar orbital moments ($L_x,L_y$) have antiferro order. Here $U=8$, $J_{\rm H}=U/6$, $t_4=-1.0$.} 
\label{fig1}
\end{figure}

Starting with AFO + FM (planar) order (Fig. \ref{fig1}) involving dominantly $yz,xz$ orbitals at $\epsilon_{xy} \sim +1$, we find this state (a) to be robustly self consistent and $m_z$ remains zero for all three orbitals. With decreasing $\epsilon_{xy}$, small $xy$ orbital moments emerge with planar AFM order as shown in (b). Finally, at $\epsilon_{xy} \sim -1$, planar AFM state with dominantly $xy$ moments is obtained (c). 

We next consider the stability of the AFO order (Fig. \ref{fig1}(a)) with respect to small magnetization perturbation. For FM-structured $\delta m_z$ (same sign on both sublattices), we find a magnetic instability which eventually leads to the orbital entangled FM $(z)$ state with $n_{yz}=n_{xz}$ as depicted in (d). First, the $m_z$ moments increase, and then the staggered orbital order decreases and eventually disappears as the orbital entangled order is formed self consistently. This finding that the staggered orbital order is unstable towards the orbital entangled order is a significant new result of our generalized self-consistent approach. With decreasing $\epsilon_{xy}$, the $xy$ orbital also gets incorporated in the entangled order (e). When a small AFM-structured magnetization perturbation $\delta m_z$ is introduced in the planar AFM order (c) with dominantly $xy$ moments at $\epsilon_{xy} \sim -1$, we find a weak instability towards axial AFM ($z$) order (f). 


\begin{figure}
\vspace*{0mm}
\hspace*{0mm}
\psfig{figure=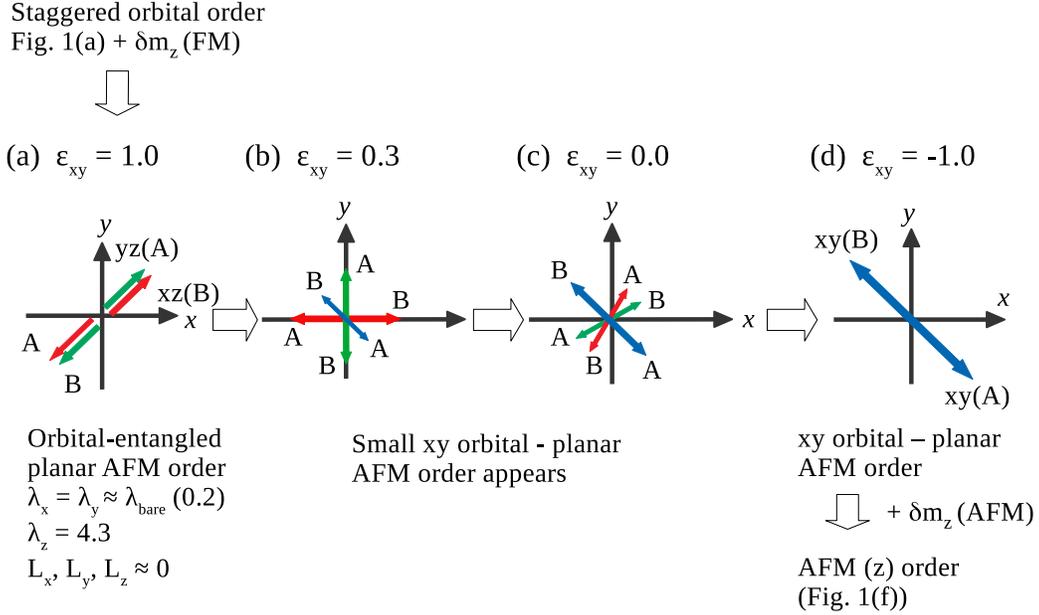,angle=0,width=140mm}
\caption{Evolution of the orbital entangled planar AFM state with the crystal field term $\epsilon_{xy}$. Here $U=12$, $J_{\rm H}=U/10$, $t_4=-0.8$.} 
\label{fig2}
\end{figure}

For the case (ii) with realistic value ($U=12|t_1|=3$ eV) for $3d$ transition metals, we find the emergence of an orbital entangled planar AFM order in our self consistent calculation. Starting with the same staggered orbital order as in Fig. \ref{fig1}(a) with the small $\delta m_z$ perturbation, we find that $m_z$ decreases with iterations, and opposite spin density in the minority orbital grows, leading to a robust orbital entangled planar AFM order as shown in Fig. \ref{fig2}(a). 

Since the effective FM interaction for staggered orbital order as obtained from the strong coupling expansion is $\sim t_4^2 J_{\rm H}/U^2$, the FM interaction is weakened for reduced $t_4,J_{\rm H}$ and increased $U$ values in case (ii), thus tilting the competition in favor of the orbital entangled AFM order. This frustration due to competing magnetic interactions accounts for the extremely low net magnetic interaction, magnon energy, and the N\'{e}el temperature. 

The orbital entangled AFM order is characterized by vanishing total spin moment on each site as the $yz$ and $xz$ moments are oppositely oriented. Also, the orbital moments $L_x,L_y,L_z$ are found to be very small. With no net local magnetic moment, this magnetically inactive state will be immune to magnetic excitations by conventional probes such as neutron scattering. Significantly, the $z$ component of Coulomb renormalized SOC is found to be huge ($\lambda_z \sim 4$), reflecting strong spin-orbital correlation induced by the Coulomb interaction terms.

Evolution of the orbital entangled AFM order with decreasing crystal field term $\epsilon_{xy}$ is shown in Fig. \ref{fig2}. We find this order to be robust and unchanged till $\epsilon_{xy}=0.5$. Thus, there is a broad crystal field (equivalently, pressure) range in which the orbital entangled AFM order is stabilized, with no change in the key characteristic values as given in Fig. \ref{fig2}(a). Below $\epsilon_{xy}=0.5$, the $xy$ orbital density starts developing, which is consistent with the energy of the $xy$ orbiton (involving $xy$ and $yz/xz$ orbitals) crossing zero as $\epsilon_{xy}\rightarrow 0.5$, as discussed later. Negative energy of the $xy$ orbiton mode would correspond to spontaneous charge excitation from $yz,xz$ orbitals to the $xy$ orbital. The planar AFM order obtained with dominantly $xy$ moments for $\epsilon_{xy}=-1.0$ as shown in Fig. \ref{fig2}(d) is weakly unstable towards the axial ($z$) AFM order when a weak AFM structured perturbation $\delta m_z$ is introduced, reflecting extremely weak easy-axis anisotropy.

\begin{figure}
\vspace*{0mm}
\hspace*{0mm}
\psfig{figure=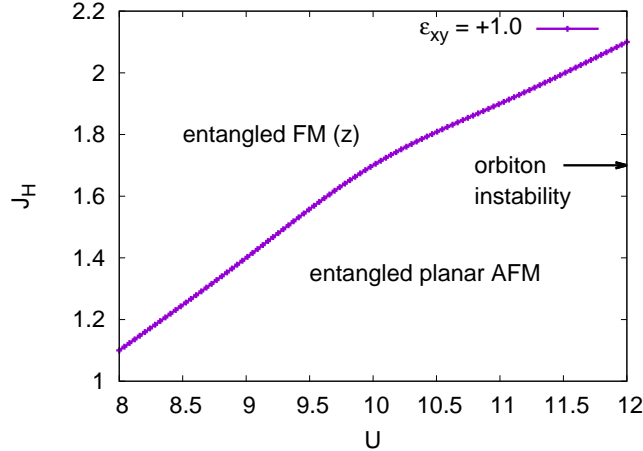,angle=0,width=90mm}
\caption{Phase boundary between the entangled AFM and entangled FM states for $\epsilon_{xy}=+1.0$.} 
\label{fig3}
\end{figure}

Fig. \ref{fig3} shows the $U$-$J_{\rm H}$ phase boundary between the orbital entangled FM and AFM orders. Here $t_4$=$-1.0$ and we have taken $\epsilon_{xy}$=$+1.0$ so that only $yz,xz$ orbitals are involved in the entangled states. The entangled planar AFM order was taken as the initial configuration in the self consistent calculation, and a small perturbation $\delta m_z$ was introduced in the $yz,xz$ moments. Increasing/decreasing $m_z$ with iterations was taken to imply approach towards the FM $(z)$ / planar AFM order. The entangled planar AFM order is seen to be stabilized with increasing $U$ and decreasing $J_{\rm H}$, which is consistent with the competition between effective FM ($ \sim t_4^2 J_{\rm H}/U^2$) and AFM ($ \sim t_4^2 /U$) inter-site interactions between the local $J=3/2$ isospin moments involving the entangled states constituted by the $yz,xz$ orbitals.  

\begin{figure}
\vspace*{0mm}
\hspace*{0mm}
\psfig{figure=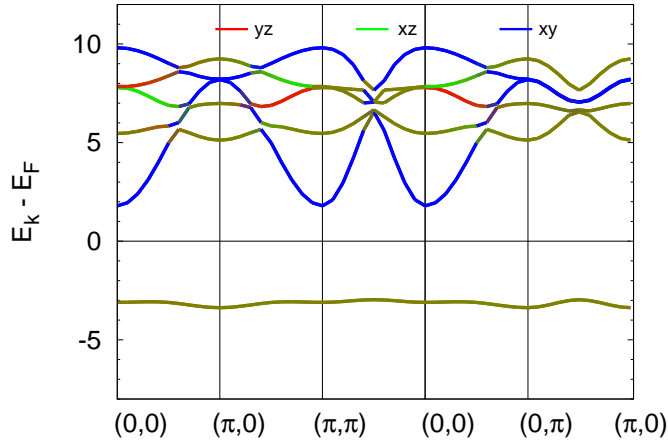,angle=0,width=90mm}
\caption{Orbital resolved electronic band structure in the orbital entangled AFM state, calculated for the case (ii) parameter values: $U=12$, $J_{\rm H}=U/10$, and $t_4=-0.8$, with $\epsilon_{xy}=+1.0$.} 
\label{fig4}
\end{figure}

We have also investigated the full range of SOC values between 0 and 0.2 (= 50 meV) in order to highlight the competition between SOC and Hund's coupling. While strong SOC favours the orbital entangled planar AFM order with locally opposite $yz,xz$ moments (Fig. \ref{fig2}a), strong Hund's coupling will favour locally parallel $yz,xz$ moments which will tend to destabilize the entangled orbital order. For $\epsilon_{xy}=+1$, $U=12$ and $J_{\rm H}=U/6$, we find the critical bare SOC value $\lambda^* \sim 0.13$. For $\lambda > \lambda^*$, the orbital entangled planar AFM order is stable, while for $\lambda < \lambda^*$ this order is destabilized in favour of the orbital staggered $(n_{yz} \ne n_{xz})$ planar FM order (Fig. \ref{fig1}a). The orbital entangled planar AFM order is therefore relevant for the $\rm Sr_2 VO_4$ compound at ambient pressure for the realistic bare SOC value $\lambda = 0.2 > \lambda^*$. 

The orbital entangled FM ($z$) order, which is stabilized for relatively lower $U$ and higher $J_{\rm H}$ values as seen in Fig. \ref{fig3}, similarly reverts back to orbital staggered FM ($z$) order when SOC is reset to zero. As the $yz,xz$ moments are parallel in both orders, instead of the Hund's coupling the main competition here is between the Coulomb interaction contributions in the orbital entangled order (spin-orbital field induced by SOC) and orbital staggered order (staggered field induced by $U''$). 

Fig. \ref{fig4} shows the calculated electronic band structure in the entangled AFM state. The degenerate bands below the Fermi energy originate from the $J=3/2$ sector entangled states $[|yz,\sigma\rangle \pm i |xz,\sigma\rangle]/\sqrt{2}$, where $\sigma=\uparrow/\downarrow$ for the sign $+/-$ corresponding to the $m_J=\pm 3/2$ states. The dominantly $xy$ orbital bands (blue) evolve from the $J=1/2$, $m_J=\pm 1/2$ states. The dominantly $yz/xz$ bands near the top correspond to $J=3/2$, $m_J=\mp 3/2$ states, while the degenerate bands near energy 5 correspond to $J=3/2$, $m_J=\pm 1/2$ states. The pure SOC eigenstates are strongly modified by the strong crystal field term $\epsilon_{xy}$. 

\section{Collective excitations}

\begin{figure}
\vspace*{0mm}
\hspace*{0mm}
\psfig{figure=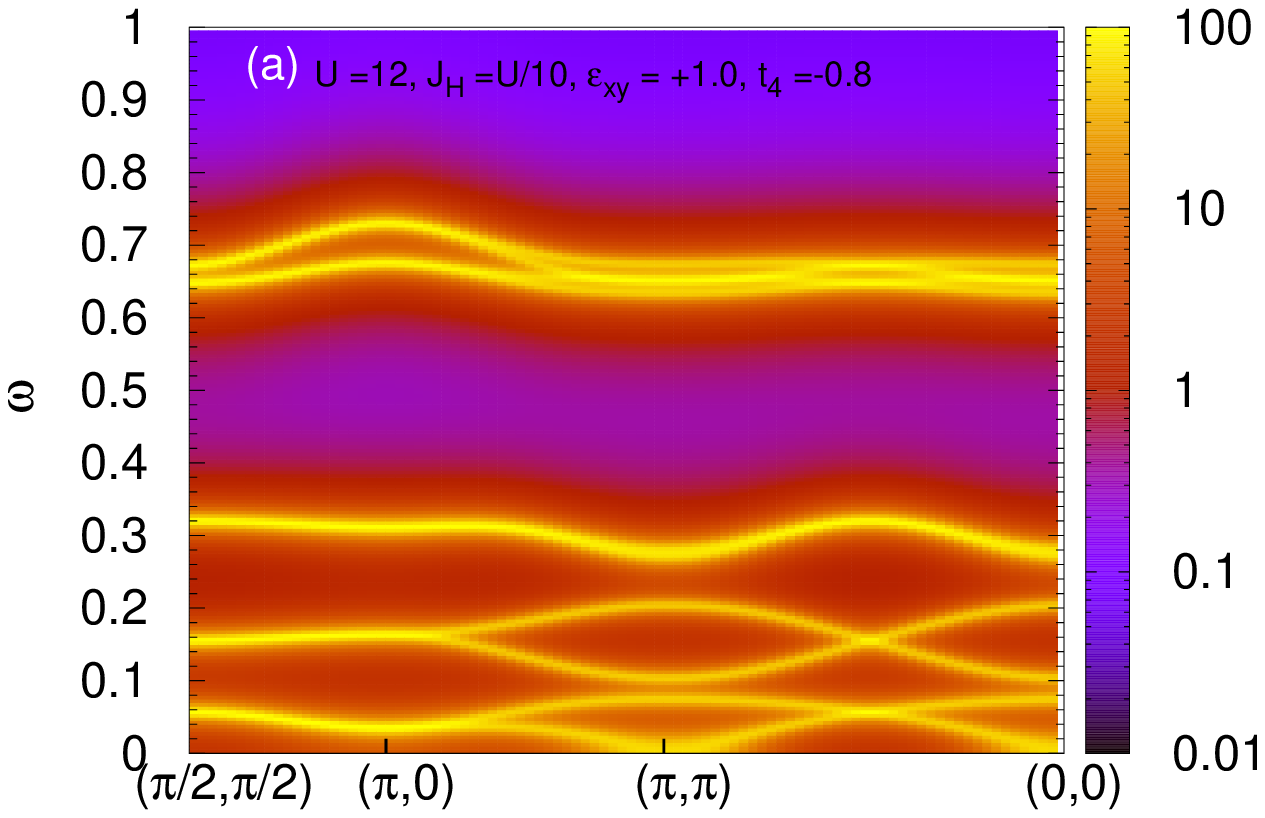,angle=0,width=53mm}
\psfig{figure=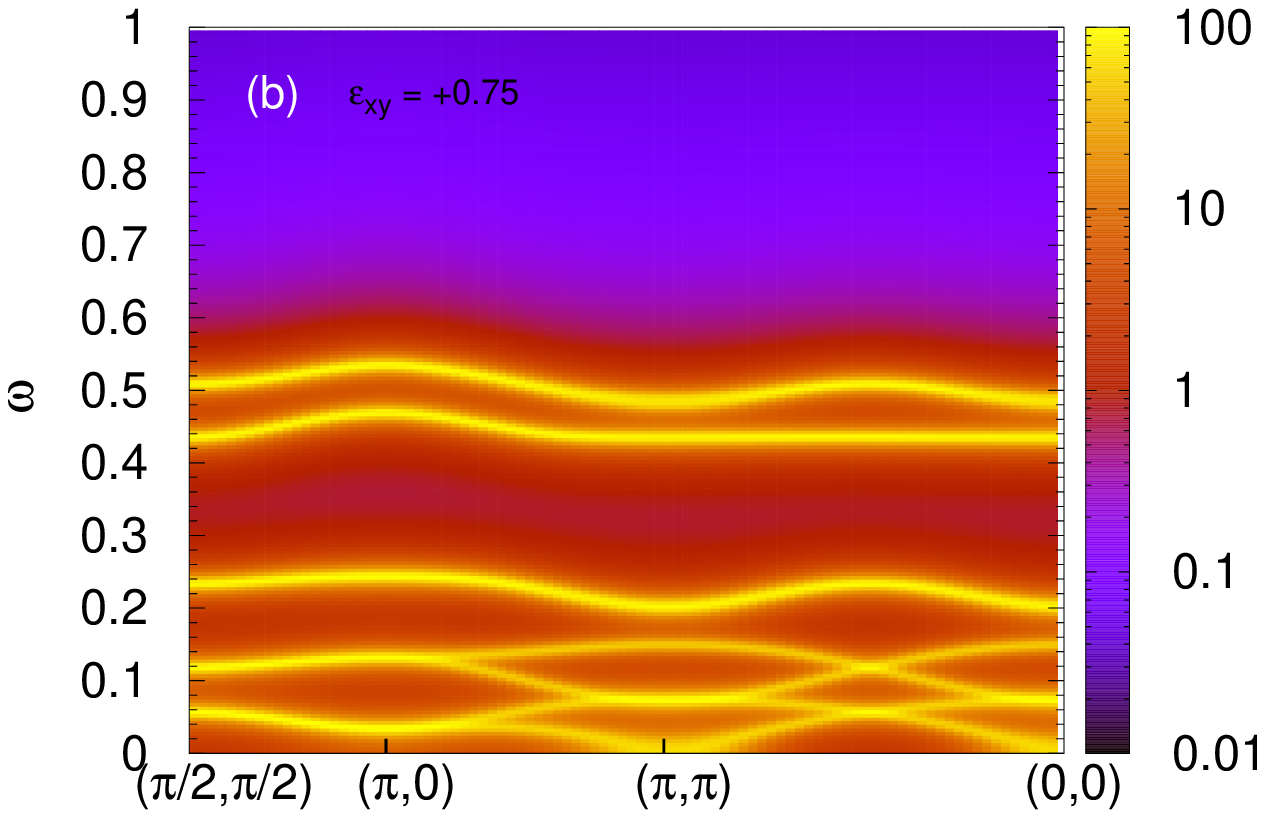,angle=0,width=53mm}
\psfig{figure=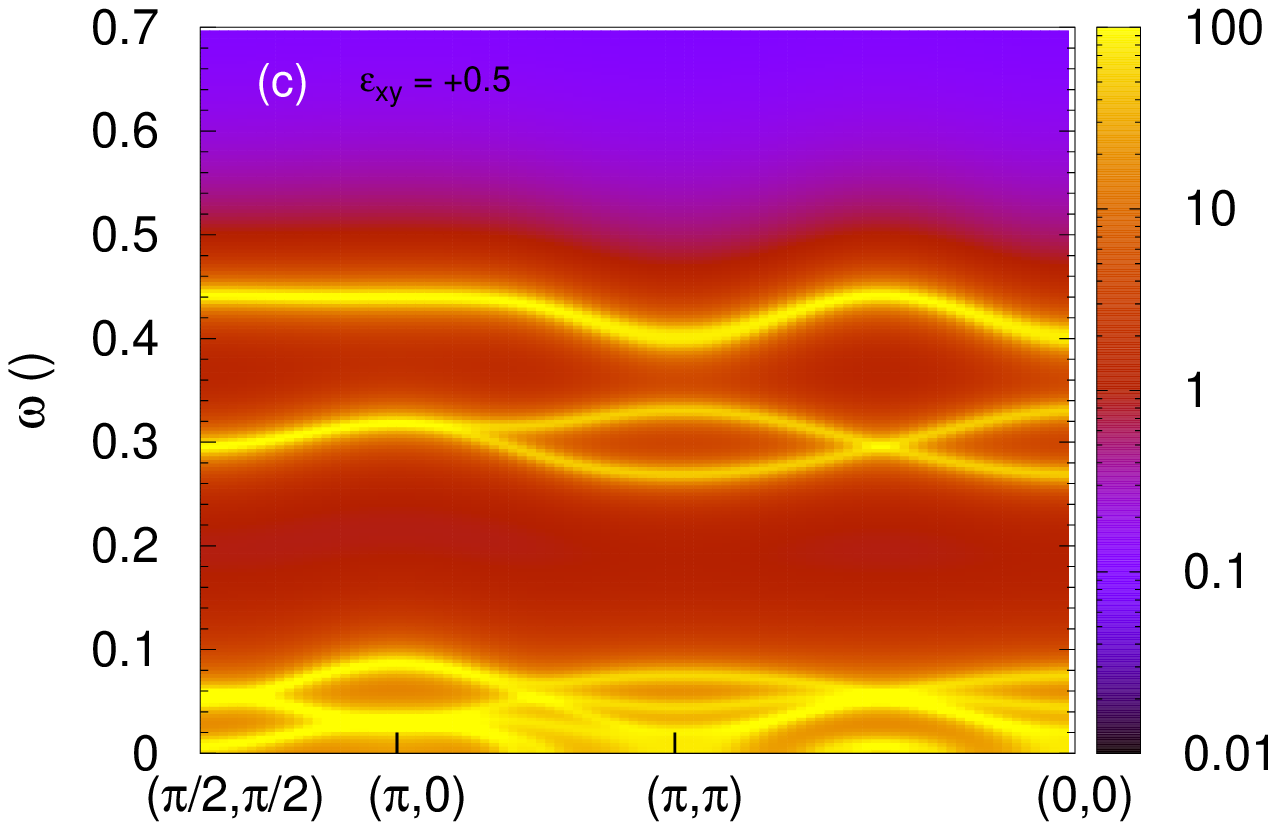,angle=0,width=53mm}
\caption{The various collective excitation modes in the entangled AFM state as obtained from the calculated generalized fluctuation propagator for: (a) $\epsilon_{xy}=1.0$, (b) 0.75, and (c) 0.5, with the same parameter set as in Fig. \ref{fig4}. The $xy$ orbiton mode energy is seen to decrease rapidly to zero as $\epsilon_{xy}$ approaches 0.5, below which $xy$ orbital density starts developing.} 
\label{fig5}
\end{figure}

Fig. \ref{fig5} shows the various collective excitation modes in the orbital entangled AFM state as obtained from the calculated spectral function (Eq. \ref{spectral}) of the generalized fluctuation propagator. In Fig. \ref{fig5}(a), the lowest mode is the extremely low energy magnon mode showing the splitting near $(\pi,\pi)$ and $(0,0)$ corresponding to in-plane and out-of-plane isospin fluctuation modes. The magnon mode involves particle-hole excitations between the $J=3/2$, $m_J=\pm 3/2$ states. The second and third modes are the spin-orbiton and orbiton modes, respectively, involving particle-hole excitations between the $m_J=\pm 3/2$ (hole) and $m_J=\pm 1/2$ (particle) states of the $J=3/2$ sector. Finally, the highest-energy modes are orbiton (same-spin) and spin-orbiton (spin-flip) modes involving particle-hole excitations between the dominantly $xy$-like $J=1/2$, $m_J=\pm 1/2$ (particle) states and the $J=3/2$, $m_J=\pm 3/2$ (hole) states. The $xy$ orbiton mode energy initially decreases slowly (a,b) as $\epsilon_{xy}$ decreases from 1.0 to 0.75, and then drops sharply to nearly zero as $\epsilon_{xy}\rightarrow 0.5$ (c). For $\epsilon_{xy} = 0.75$ [Fig. \ref{fig5}(b)], the energy scale of this pair of modes ($\sim 0.5 \times 250$ meV = 125 meV) is in good agreement with the INS study.\cite{zhou_PRB_2010}

\begin{figure}
\vspace*{0mm}
\hspace*{0mm}
\psfig{figure=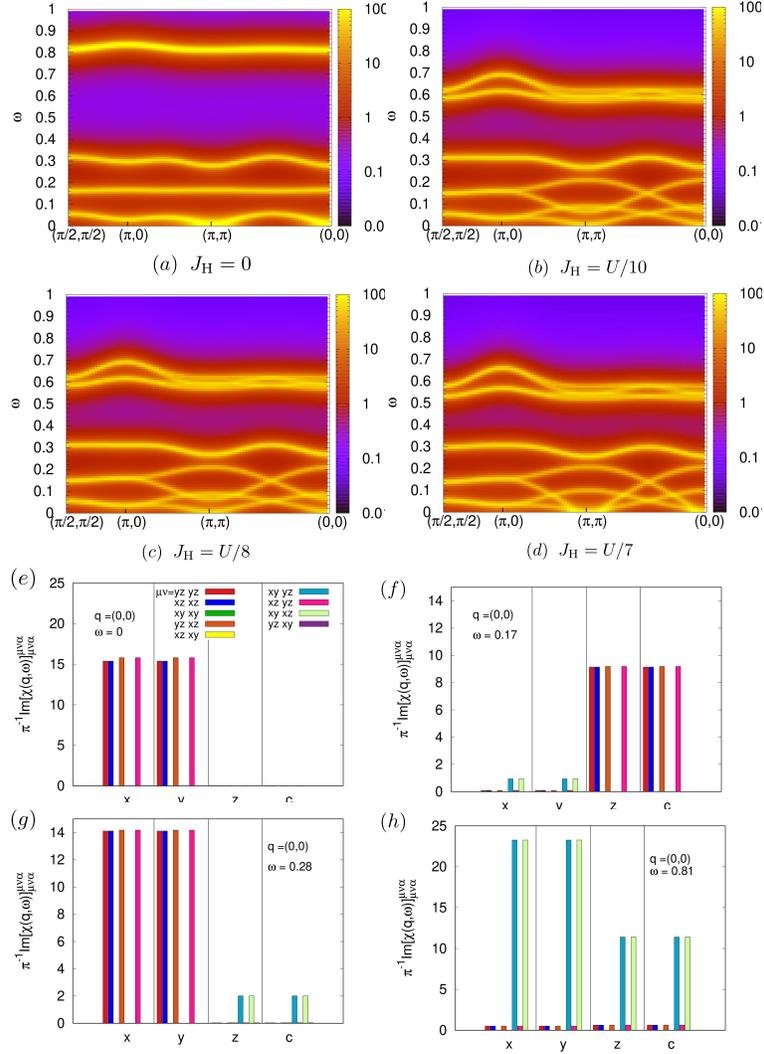,angle=0,width=100mm} 
\caption{(a)-(d) Evolution of the collective excitation energies with increasing $J_{\rm H}$ values, showing that the orbiton lower branch energy at ${\bf q}=(0,0)/(\pi,\pi)$ vanishes for $J_{\rm H}\sim U/7$. (e)-(h) Orbital-pair-basis resolved contributions to the spectral function at ${\bf q}=(0,0)/(\pi,\pi)$ for the different collective excitation modes shown in (a) for $J_{\rm H}=0$. Other parameters are same as in Fig. \ref{fig4}.} 
\label{fig6}
\end{figure}

In Fig. \ref{fig5}(a), we consider the splitting of the magnon mode at ${\bf q}=(\pi,\pi)$ or $(0,0)$ into the gapless (in-plane) and gapped (out-of-plane) isospin fluctuation modes. We find that when $J_{\rm H}$ is set to zero, the two modes become degenerate and gapless at these wave vectors (Fig. \ref{fig6}(a)), implying no true anisotropy due to SOC alone. This feature is exactly similar as in the $\rm Sr_2 IrO_4$ compound where the easy-plane anisotropy was identified as arising due to the $J_{\rm H}$ induced spin-rotation-symmetry breaking in the pseudo spin-orbital basis.\cite{mohapatra_JPCM_2021,mohapatra_pseudo_JPCM_2021} With one electron in $\rm Sr_2VO_4$ and one hole in $\rm Sr_2IrO_4$, the above similarity reflects the particle-hole symmetry at play.  

\begin{figure}
\vspace*{-5mm}
\hspace*{0mm}
\psfig{figure=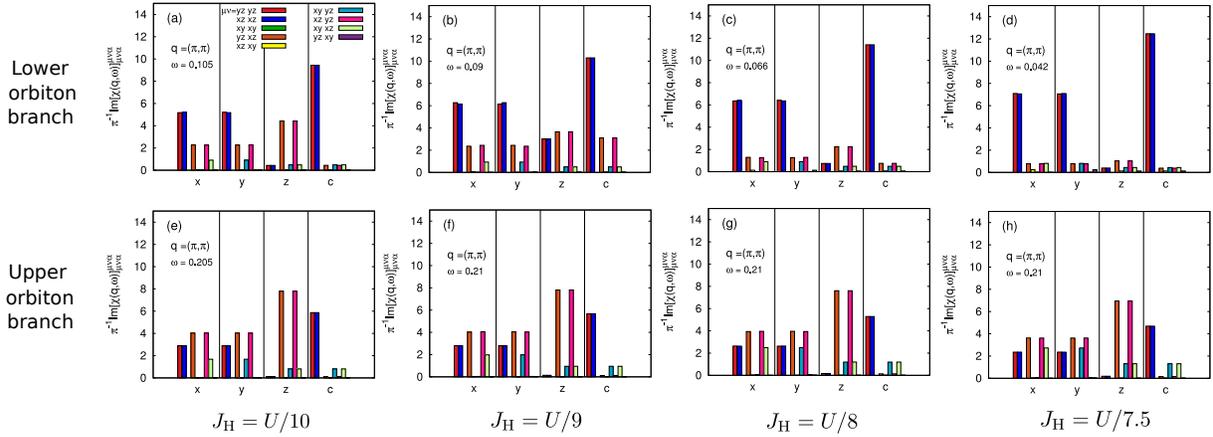,angle=0,width=160mm}
\caption{Evolution of the basis resolved contributions to the spectral function at ${\bf q}=(\pi,\pi)$ for the lower and upper branches of the orbiton mode with increasing $J_{\rm H}$ values.} 
\label{fig7}
\end{figure}

For the spin-orbiton mode (second lowest in Fig. \ref{fig5}(a)), we find that the minimum energy of the lower branch at ${\bf q}=(0,0)$/$(\pi,\pi)$ decreases with increasing $J_{\rm H}$ and vanishes at the critical value $J_{\rm H}^* \sim U/7$, as seen in Fig. \ref{fig6}. Subsequently, this mode becomes negative-energy mode for $J_{\rm H}>J_{\rm H}^*$, indicating instability due to long wavelength $yz/xz$ orbital fluctuations. This instability is shown by the arrow in Fig. \ref{fig3}. We have also analyzed the different spectral function contributions in the orbital-pair basis for this mode at ${\bf q}=(0,0)/(\pi,\pi)$. We find that with increasing $J_{\rm H}$ the entanglement character (measured by the $\mu,\nu=yz,xz$ contribution) progressively decreases (Fig. \ref{fig7}) and becomes negligible as $J_{\rm H} \rightarrow J_{\rm H}^*$. Therefore, the instability is due to small-${\bf q}$ (long wavelength) fluctuation modes of nearly pure spin-charge character and short-range entangled AFM order is expected to survive. 

\begin{figure}
\vspace*{0mm}
\hspace*{0mm}
\psfig{figure=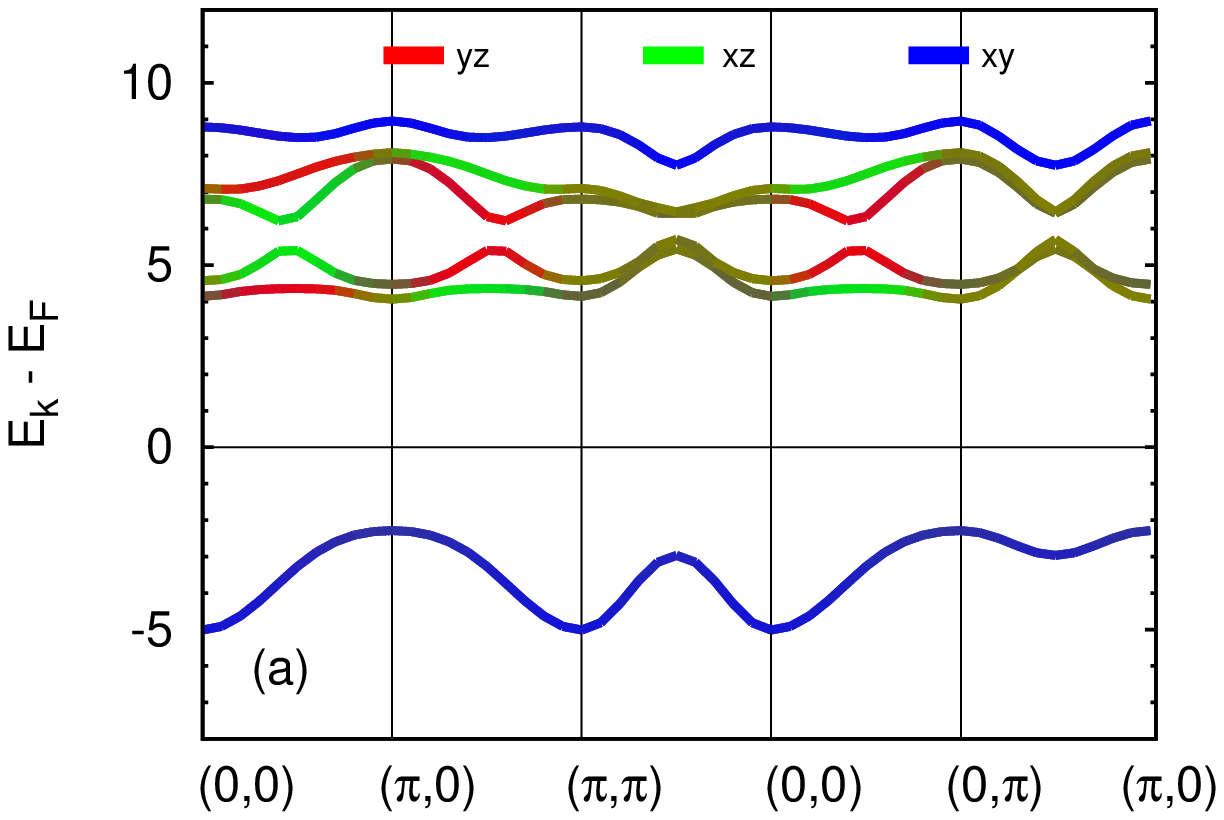,angle=0,width=75mm}
\psfig{figure=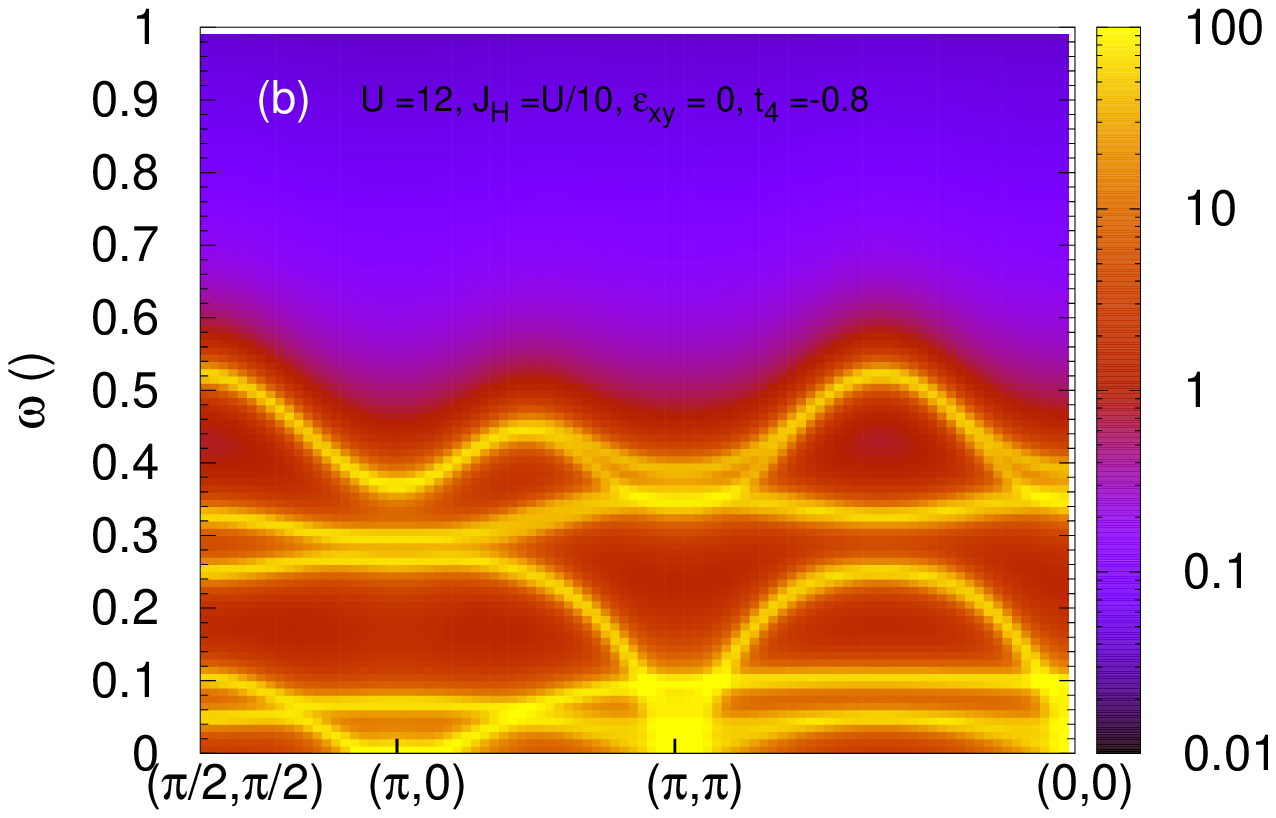,angle=0,width=75mm}
\caption{(a) Orbital resolved electronic band structure in the orbital entangled AFM state, and (b) the various collective excitation modes, calculated for the crystal field term $\epsilon_{xy}=0$, with the same parameter set as in Fig. \ref{fig4}.} 
\label{fig8}
\end{figure}

For $\epsilon_{xy}=0$ where the planar AFM order involves dominantly $xy$ moments, the electronic band structure and the various collective excitation modes are shown in Fig. \ref{fig8}. The extremely low-energy orbiton modes are seen to be present in this case also, although the two pairs of orbiton modes here involve particle-hole excitations between the dominantly $xy$ and $yz,xz$ bands corresponding to the entangled $J=3/2$, $m_J=\pm 1/2$ (hole) and $J=3/2$, $m_J=\pm 3/2$ or $J=1/2$, $m_J=\pm 3/2$ (particle) states. The magnon modes involving dominantly $xy$ orbital are split into the low-energy part near zone center and high-energy part ($\omega \sim 0.5$) near the zone boundary.

\section{Spin-orbit-distortion exciton}

\begin{figure}
\vspace*{0mm}
\hspace*{0mm}
\psfig{figure=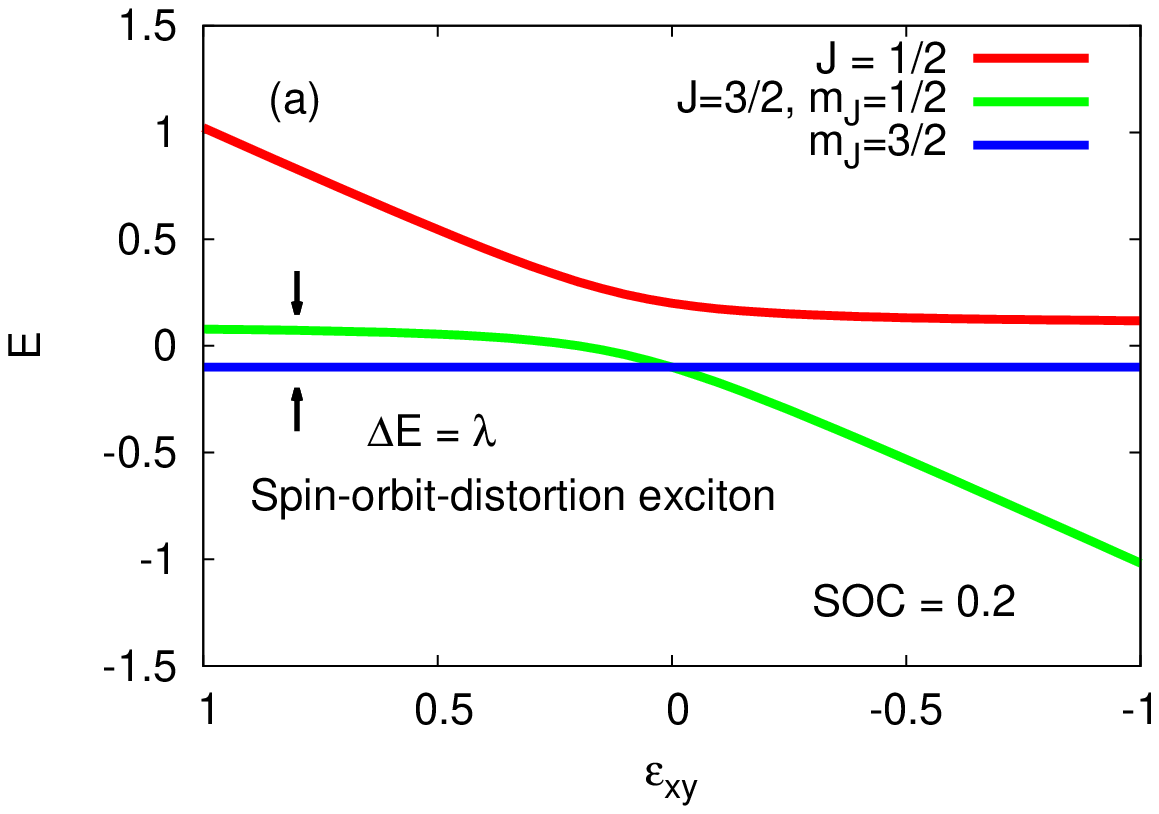,angle=0,width=80mm} \hspace{-10mm}
\psfig{figure=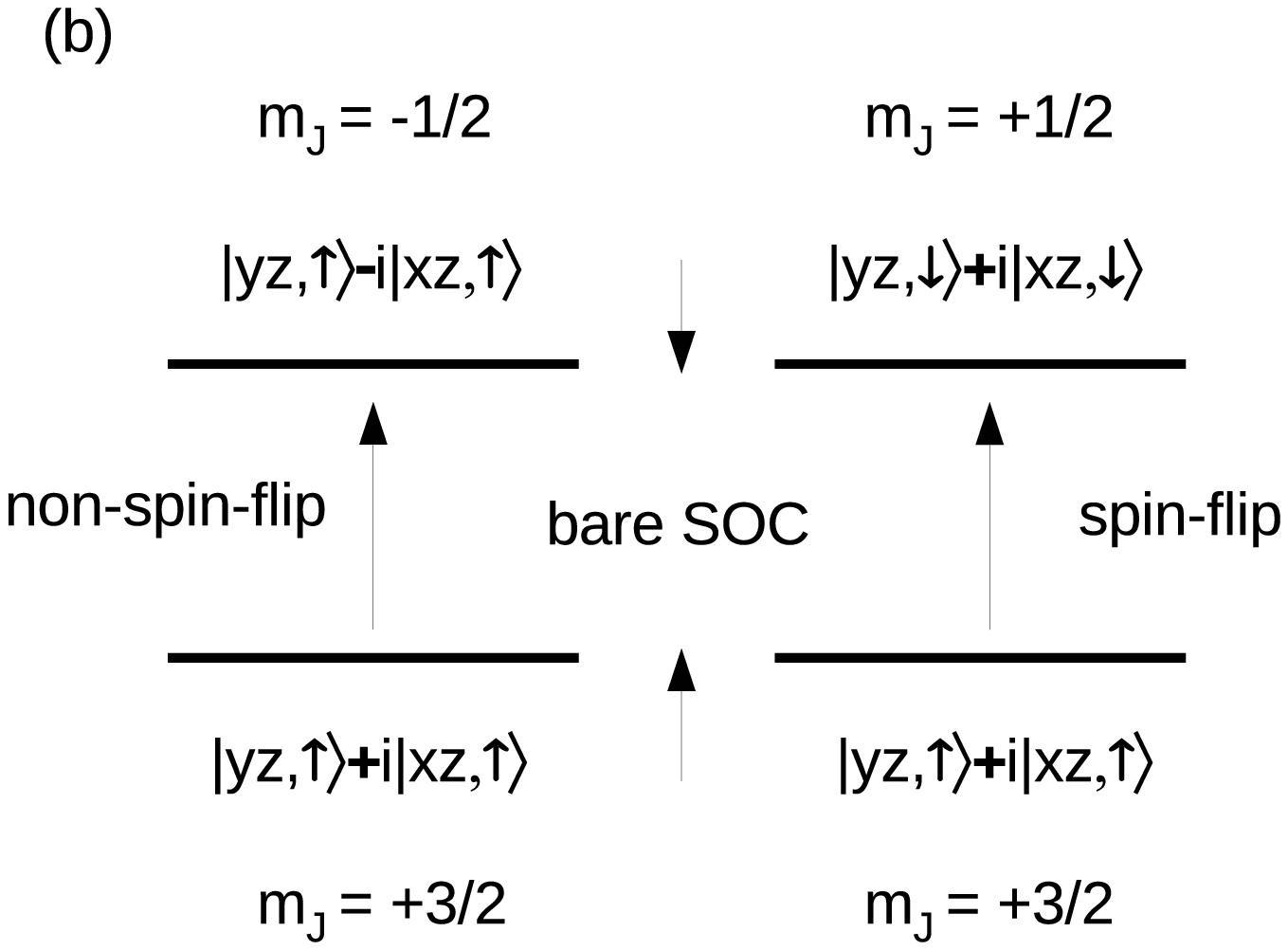,angle=0,width=90mm}
\caption{(a) Evolution of the energy eigenvalues of ${\cal H}_{\rm SOC} + {\cal H}_{\rm cf}$ with crystal field. (b) Excitations between the $J=3/2$ sector states (for $\epsilon_{xy} \gg \lambda$), showing the non-spin-flip and spin-flip cases corresponding to the orbiton and spin-orbiton modes in Fig. \ref{fig5}.} 
\label{fig9}
\end{figure}

To understand the origin of the low-energy orbiton modes in Fig. \ref{fig5}, it is instructive to consider the evolution of the energy eigenvalues of the ${\cal H}_{\rm SOC} + {\cal H}_{\rm cf}$ Hamiltonian with the crystal field term $\epsilon_{xy}$ as shown in Fig. \ref{fig9}. The spin-orbital entangled $J=3/2$ sector states are split by the tetragonal distortion induced crystal field, and the splitting $\Delta E = \lambda$ (the bare SOC value) as $\epsilon_{xy}/\lambda \rightarrow \infty$. Thus, the lowest energy particle-hole excitations are between the $J=3/2$ sector states, involving dominantly $yz,xz$ character at large positive $\epsilon_{xy}$, with extremely low excitation energy for the $3d$ transition metal compounds.

We will refer to the collective (orbiton mode) excitations arising from the $J=3/2$ sector particle-hole excitations as spin-orbit-distortion excitons. The energy separation between the bands originating from the $m_J=\pm 3/2,\pm 1/2$ of the $J=3/2$ sector are strongly enhanced by the various Coulomb interaction contributions (Fig. \ref{fig4}). However, due to the usual resonant scattering process in the random phase approximation (RPA), the excitation energies are lowered down to nearly the bare exciton energies. Thus, the two orbiton modes (second and third lowest energy) in Fig. \ref{fig5}(a) are spin-orbit-distortion exciton modes, and the extremely low excitation energies are nearly independent of the crystal field term $\epsilon_{xy}$ and of the order of the bare SOC value. This leads to the surprising conclusion that the measured $\chi-T$ and $\rho-T$ anomalies in $\rm Sr_2 VO_4$ are due to thermal excitation of modes for which the characteristic energy is the extremely low bare SOC value for $3d$ elements.  

\section{Discussion}

The presence of multiple low-energy orbiton modes over a broad crystal field range highlights the rich orbital physics in the $\rm Sr_2VO_4$ compound. To illustrate the importance of the low-energy orbiton modes, we consider the $xy$ orbiton modes which rapidly lower in energy as $\epsilon_{xy}\rightarrow 0.5$ (Fig. \ref{fig5}). Thermal excitation of these modes at finite temperature will result in electron transfer from the magnetically inactive entangled state with no local magnetic moment (Fig. \ref{fig2}) to the magnetically active $xy$ orbital, resulting in finite magnetic moment and contribution to magnetic susceptibility.

The behaviour of the collective excitation mode energies (Fig. \ref{fig5}) with decreasing crystal field term $\epsilon_{xy}$ shows: (i) extremely low and nearly constant energy of magnon excitations, (ii) low energy orbiton modes involving $yz,xz$ orbitals, and (iii) rapidly decreasing energy of the $xy$ orbiton modes to zero as $\epsilon_{xy}\rightarrow 0.5$. These features are in striking similarity to the behaviour of the three transition temperatures with pressure in the $0-5$ GPa range obtained in the high-pressure study from anomalies in resistivity and susceptibility measurements in $\rm Sr_2VO_4$.\cite{yamauchi_PRL_2019} The magnetic transition at 8K obtained from susceptibility measurements corresponds to the N\'{e}el temperature, which supports the picture of thermal excitation of the extremely low energy magnons ($\sim 10$ meV) in the orbitally entangled AFM state as driving the magnetic transition.  

\section{Conclusions}

Our main findings are summarized below. Staggered orbital order is unstable towards orbital entangled (FM or AFM) orders depending on the $U$ and $J_{\rm H}$ values. When SOC is reset to zero after the entangled order is self consistently formed, staggered orbital order reappears, confirming that the orbital entangled states are induced by SOC. The obtained phase boundary in $U-J_{\rm H}$ space between entangled FM and AFM orders confirms the qualitative analysis that for fixed $J_{\rm H}$ the entangled AFM order is stabilized on the higher $U$ side, which lies within the realistic range for $3d$ transition metal elements. The orbiton mode shows long-wavelength instability with increasing $J_{\rm H}$, but short range entangled AFM order appears robust. 

In the entangled AFM order, frustration due to competing magnetic interactions accounts for the extremely low net magnetic interaction, magnon energy, and therefore the N\'{e}el temperature. The easy-plane anisotropy vanishes when Hund's coupling is set to zero, indicating no true anisotropy due to SOC alone, which is similar to the case of $\rm Sr_2 IrO_4$. The extremely low ($\sim 10$ meV) magnon energy scale, the low-energy $yz/xz$ orbiton modes having energy of the order of bare SOC value over a broad crystal field range, the $\sim$ 125 meV energy of the $xy$ orbiton mode at ambient pressure, and the behaviour of the $xy$ orbiton energy with crystal field are all in good agreement with experiments. 


\begin{acknowledgments}
DKS was supported through start-up research grant SRG/2020/002144 funded by DST-SERB.
\end{acknowledgments}

\end{document}